# Infrared Supercontinuum Generation in Multiple Quantum Well Nanostructures under Electromagnetically Induced Transparency

Nitu Borgohain[a*], Milivoj Belić[b] and S. Konar[a]
[a] Department of Physics, Birla Institute of Technology, Mesra-835215, Ranchi, India,
[b] Department of Physics, Texas A&M University, PO Box 23874, Doha, Qatar.

*E-mail: nituborgohain.ism@gmail.com

**Abstract**

Mid-infrared spectral broadening is of great scientific and technological interest, which till date is mainly achieved using non-silica glass fibers, primarily made of tellurite, fluoride and chalcogenide glasses. We investigate broadband mid-infrared supercontinuum generation at very low power in semiconductor multiple quantum well (MQW) systems facilitated by electromagnetically induced transparency. 100 femto-seconds pulses of peak power close to a Watt have been launched in the electromagnetically induced transparency window of a 30 period 1.374 $\mu m$ long MQW system. Broadband supercontinuum spectra, attributed to self phase modulation and modulation instability, is achievable at the end of the MQW system. The central part of the spectra is dominated by several dips and the far infra-red part of the spectra is more broadened in comparison to the infra-red portion. Key advantage of the proposed scheme is that the supercontinuum source could be easily integrated with other semiconductor devices.

Recently optical supercontinuum (SC) generation has drawn tremendous attention[1-10] due to several important applications such as optical coherence tomography[5], optical metrology[6], spectroscopy, optical frequency comb generation[7,8], and wavelength division multiplexing[9]. Since discovery[1], SC generation has been studied in different nonlinear media including optical fibers[2-8], silicon photonic nanowires[10], chalcogenide waveguides[11-12] and silica waveguides[13,14]. Though the SC generation has been experimentally achieved in different media, the photonic crystal fibers (PCFs) have emerged as the most popular nonlinear media for SC generation due to the feasibility of dispersion and nonlinearity engineering[2,3]. The discovery of nonsilica PCFs, characterized by large optical nonlinearity, has further enhanced their popularity as a nonlinear medium for successful SC generation[15,16].

SC generation is characterized by dramatic spectral broadening of an optical field which occurs when an intense narrowband light pulse propagates through a nonlinear medium[2-4]. The spectral broadening is contributed by a host of nonlinear optical process such as self-phase modulation, cross-phase modulation, modulation instability, soliton fission, Raman scattering, dispersive wave generation, four wave mixing, self-steepening[2-4] etc. These nonlinear processes are governed by pulse duration, wavelength and peak power of the pump pulse, whereas the group velocity dispersion (GVD) and its higher order terms at the pumping wavelength play a vital role in determining the quality of the continuum and its shape. Usually, in a PCF the SC spectra are generated by pumping nanosecond, picosecond or femtosecond[2-4] pulses whose wavelength is in the anomalous dispersion regime that is close to the zero dispersion point. However, in PCFs the SC can be generated in the normal dispersion region too[17] where the spectral broadening is dominated through self-phase modulation, Raman scattering and four wave mixing[2,3]. PCFs with large effective nonlinearity require very low threshold power to generate the wideband SC, on the other hand low and uniform dispersion enables four-wave matching leading to wideband flat spectra. Thus, large effective optical nonlinearity and low uniform dispersion profile is the key to broadband flat SC spectra. Recently Lau et al.[18] have reported SC generation in a silicon nano-waveguide by employing sech pulses of peak power as low as 60W. Identification of appropriate highly nonlinear material is central issue to SC generation at low power. Since in comparison to PCFs, giant optical nonlinearity could be experienced in semiconductor quantum well (QW) nanostructures under appropriate experimental configuration, it may be possible to achieve SC generation in QWs at much lower power level in comparison to PCFs. Moreover, they can be easily integrated with other semiconductor devices.

In recent years, quantum coherence and interference effects in semiconductor nanostructures, particularly in QWs, have received tremendous attention due to the widespread use of semiconductor components in optical computing, optical communication and quantum information processing. Quantum coherent phenomena such as gain without inversion (GWI), coherent population oscillations, electromagnetically induced transparency (EIT), and slow light propagation have been explored both theoretically and experimentally[19-24] in QWs. Frogley et al.[21] were the first to experimentally demonstrate GWI in a specifically designed quantum structure made of three InGaAs/AlInAs quantum wells. Success of this experiment paved the way for the foundation of several active devices such as lasers that do not require population inversion, and semiconductor based 'slow light' devices[21,25,26] that directly leads to applications in optical data storage and computing. Chang et al.[27] investigated the slow light phenomenon in a quantum well waveguide using EIT. They pointed out that long term electron-spin life time in [110] quantum well and the strain-induced shift of the light-hole-like excitonic transition energy below those of the heavy-hole-like continuum states can enhance the performance of slow light. Palinginis et al.[28] experimentally realized slow light propagation via coherent population oscillation in a GaAs quantum well waveguide. Yang et al.[29] proposed a theoretical scheme to realize four-wave mixing via electron spin coherence in a waveguide. Their results displayed that electromagnetically induced absorption and superluminal propagation could be realized in the waveguide at room temperatures. All optical switching, four-wave mixing and slow optical solitons in QWs have been investigated successfully using EIT. The implementation of EIT in semiconductor-based devices is very attractive from a viewpoint of applications. Devices based on intersubband transitions in semiconductor QW structures have many inherent advantages over other systems, such as large electric dipole moments due to the small effective electron mass consequently giant nonlinear optical coefficients, and a great flexibility in device design since their transition energies, dipole moment and symmetries can be engineered as desired by choosing appropriate structure dimensions and materials. Based on quantum coherence and interference effects Kerr nonlinearity can be enhanced enormously, while linear absorption and even two-photon absorption can be suppressed. Particularly, facilitated by EIT much larger optical nonlinearity can be engineered in QWs in comparison to that of PCFs and other nonlinear media, thus, it is prudent to examine SC generation in semiconductor QWs. Hence, the main thrust of this communication is to theoretically examine the possibility of generation of SC in quantum well nanostructures utilizing giant optical nonlinearity created under EIT. Therefore, in the present communication we first identify large Ker nonlinearity in a ladder type three level semiconductor multiple quantum well (MQW) system driven by a probe laser pulse and controlled by an additional coupling field. In the second step, we utilize this EIT created large nonlinearity for the generation of supercontinuum of the pump pulse at much lower peak power level in comparison to the SC generation achieved in other media as elucidated previously.

**Theoretical model and equations**

First, we consider a MQW structure with three energy levels that forms the well known 'cascade' configuration, as shown schematically in Fig. 1. This system was first demonstrated both experimentally as well as theoretically by J. F. Dynes et al. in 2005[30]. The sample consists of 30 coupled well periods, where each period consists of a 4.8 nm $In_{0.47}Ga_{0.53}As$ /0.2 nm $Al_{0.48}In_{0.52}As$ /4.8 nm $In_{0.47}Ga_{0.53}As$ coupled quantum well separated by modulation doped 36 nm $Al_{0.48}In_{0.52}As$ barriers. The lattice matched to an undoped *InP* substrate. A weak probe pulse with angular frequency $\omega_p$ and amplitude $E_p$ couples the transition between the states $|1\rangle$ and $|2\rangle$, and simultaneously a strong control laser beam with angular frequency $\omega_c$ and amplitude $E_c$ couples the transition between the states $|2\rangle$ and $|3\rangle$ of this system. The electric field of the probe pulse and control beam system can be written as,

$$\vec{E} = \hat{e}_p E_p exp\{i(k_p z - \omega_p t)\} + \hat{e}_c E_c exp\{i(k_c z - \omega_c t)\} + c.c. \tag{1}$$

where $\hat{e}_p$ and $\hat{e}_c$ are the unit vectors along the polarization direction of the probe and control field, respectively; $k_p$, $k_c$ are the wave numbers of the probe and control fields respectively.

In the Schrödinger picture with rotating wave approximation, the semi-classical Hamiltonian of the system can be written as, $\widehat{H} = \widehat{H}_0 + \widehat{H}'$, where $\widehat{H}_0$ describes the free Hamiltonian of the system in absence of any external field and $\widehat{H}'$ describes the perturbed Hamiltonian due to the interaction between MQW and the fields. In Schrödinger picture, these two parts of the Hamiltonian can be written as,

$$\widehat{H}_0 = \sum_{i=1}^{3} E_i |i\rangle\langle i|, \tag{2}$$

and $\hat{H}' = -\hbar\{\Omega_p e^{i(k_p.z - \omega_p t)}|3\rangle\langle 1| + \Omega_c e^{i(k_c.z - \omega_c t)}|3\rangle\langle 2| + h.c.\},$ (3)

where $\Omega_p$ and $\Omega_c$ are the half Rabi frequencies of the probe and control field, respectively, which are defined as $\Omega_p = \frac{(\hat{\mu}_{13}.\hat{e}_p)E_p}{2\hbar}$ and $\Omega_c = \frac{(\hat{\mu}_{23}\hat{e}_c)E_c}{2\hbar}$, while $\hat{\mu}_{mn} = e\langle m|z|n\rangle$ is the dipole matrix elements for the transition $|m\rangle \leftrightarrows |n\rangle$. To analyze the light-matter interaction process in the system we adopt the density matrix formalism[22,23,29,30], in which the evolution of the density operator '$\rho$' of the system is governed by the generalized Schrödinger equation written as:

$$\dot{\rho} = \frac{1}{i\hbar}[(\hat{H}_0 + \hat{H}'), \rho].$$ (4)

Standard procedure yields following equations of motion for the matrix elements $\rho$:

$$\dot{\tilde{\rho}}_{11} = i\Omega_p^* \tilde{\rho}_{21} - i\Omega_p \tilde{\rho}_{12},$$ (5.1)

$$\dot{\tilde{\rho}}_{22} = -\gamma_2 \tilde{\rho}_{22} + i\Omega_p \tilde{\rho}_{12} + i\Omega_c^* \tilde{\rho}_{32} - i\Omega_p^* \tilde{\rho}_{21} - i\Omega_c \tilde{\rho}_{23},$$ (5.2)

$$\dot{\tilde{\rho}}_{33} = -\gamma_3 \tilde{\rho}_{33} + i\Omega_c \tilde{\rho}_{23} + i\Omega_c^* \tilde{\rho}_{32},$$ (5.3)

$$\dot{\tilde{\rho}}_{21} = i\left(\Delta_p + i\frac{\gamma_{21}}{2}\right)\tilde{\rho}_{21} + i\Omega_p(\tilde{\rho}_{11} - \tilde{\rho}_{22}) + i\Omega_c^* \tilde{\rho}_{31},$$ (5.4)

$$\dot{\tilde{\rho}}_{32} = i\left(\Delta_c + i\frac{\gamma_{32}}{2}\right)\tilde{\rho}_{32} + i\Omega_c(\tilde{\rho}_{22} - \tilde{\rho}_{33}) - i\Omega_p^* \tilde{\rho}_{31},$$ (5.5)

$$\dot{\tilde{\rho}}_{31} = i\left(\Delta_p + \Delta_c + i\frac{\gamma_{31}}{2}\right)\tilde{\rho}_{31} + i\Omega_c \tilde{\rho}_{21} - i\Omega_p \tilde{\rho}_{32},$$ (5.6)

where $\gamma_i (i = 2,3)$ are the population decay rates which are dominated by the inelastic emission of longitudinal optical (LO) phonons. $\gamma_{ij}(i \neq j)$ represents the total coherence relaxation rates given by $\gamma_{21} = \gamma_2 + \gamma_{21}^{dph}$; $\gamma_{31} = \gamma_3 + \gamma_{31}^{dph}$ and $\gamma_{32} = \gamma_2 + \gamma_3 + \gamma_{32}^{dph}$, where $\gamma_{ij}^{dph}$ comprise the sum of quasi-elastic acoustic phonon scattering and the elastic interface Raman scattering. The detunings $\Delta_p$ and $\Delta_c$ are defined as $\Delta_p = \omega_p - \omega_{21}$ and $\Delta_c = \omega_c - \omega_{32}$, where $\omega_{ji(i \neq j)}$ are the angular frequencies of the resonant transition between states $|i\rangle \rightleftharpoons |j\rangle$. The propagation constant $\beta(\omega)$, linear and third order susceptibilities $(\chi^{(1)}$ and $\chi^{(3)})$ of the pump pulse turns out to be

$$\beta(\omega) = \frac{\omega}{c} - \kappa \frac{D_p(\omega)}{D(\omega)}, \tag{7}$$

$$\chi^{(1)} = -\frac{N|\mu_{12}|^2}{\hbar\varepsilon_0} \frac{D_p(0)}{D(0)}, \tag{8}$$

$$\chi^{(3)} = \frac{N|\mu_{12}|^4}{4\hbar^3\varepsilon_0} \frac{\left(|\Omega_c|^2+|D_p(0)|^2\right)D_p(0)}{|D(0)|^2 D(0)}, \tag{9}$$

where $D_p(\omega) = \left(\omega + \Delta_p + \Delta_c + i\frac{\gamma_{31}}{2}\right)$ and $D(\omega) = \left(\omega + \Delta_p + i\frac{\gamma_{21}}{2}\right)\left(\omega + \Delta_p + \Delta_c + i\frac{\gamma_{31}}{2}\right)$. The propagation constant $\beta(\omega)$ can be expanded in Taylor series around the central frequency of the probe field ($\omega = 0$) as,

$$\beta(\omega) = \beta(0) + \beta_1(0)\omega + \frac{1}{2}\beta_2(0)\omega^2 + \cdots, \tag{10}$$

where, $\beta_n(0) = \frac{d^n\beta(\omega)}{d\omega^n}\bigg|_{\omega=0}$. The group velocity is given by, $V_g = Re\left[\frac{1}{\beta'(0)}\right]$ and $\beta_2(0) = \frac{d^2\beta(\omega)}{d\omega^2}\bigg|_{\omega=0}$ represents the group velocity dispersion of the probe pulse leading to change of shape of the propagating pulse. The nonlinear dynamics of the pulse inside the MQW is governed by the following equation:

$$i\frac{\partial \widetilde{\Omega}_p}{\partial z} + i\beta_1(0)\frac{\partial \widetilde{\Omega}_p}{\partial t} - \frac{1}{2}\beta_2(0)\frac{\partial^2 \widetilde{\Omega}_p}{\partial t^2} + W|\widetilde{\Omega}_p|^2\widetilde{\Omega}_p = 0, \tag{11}$$

where $W = \kappa\left(\frac{|\Omega_c|^2+|D_p(0)|^2}{|D(0)|^2}\right)\frac{D_p(0)}{D(0)}$. Introducing the retarded frame $\xi = z$ and $T = t - z/V_g$, and after suitable rescaling above equation can be recasted as:

$$i\frac{\partial A}{\partial z} - \frac{1}{2}\beta_2(0)\frac{\partial^2 A}{\partial T^2} + \gamma|A|^2 A = 0, \tag{12}$$

where $A = \frac{\hbar \left(\frac{cn\varepsilon_0 S}{2}\right)^{1/2}}{\mu_{12}} \widetilde{\Omega}_p$, $\gamma = W\left(\mu_{12}/\hbar \sqrt{\frac{cn\varepsilon_0 S}{2}}\right)^2$; c, n and $\varepsilon_0$ are the velocity of light in vacuum, linear refractive index of the medium and vacuum permittivity, respectively.

**Linear and nonlinear susceptibilities**

In this section we focus on the numerical investigation of the linear and nonlinear susceptibilities of the MQW system considered in the present communication. However, we first examine the linear (first order) susceptibility of the system with the objective of achieving low absorption which may provide a useful starting point in the subsequent sections. The system parameters taken for present study are: $N = 10^{22}\ m^{-3}$, $\mu_{12} = 23.35\ e\text{Å}$, $\omega_p = 18.84 \times 10^{13} s^{-1}$, thus, $\kappa = 4.69 \times 10^{17} m^{-1} s^{-1}$; decay rates $\gamma_{21} = 0.5 \times 10^{12} s^{-1}$ and $\gamma_{31} = 1.0 \times 10^{12} s^{-1}$. To study the behaviour of real and imaginary parts of the linear susceptibility, we demonstrate the variations of $Im(\chi^{(1)})$ and $Re(\chi^{(1)})$ as a function of normalized probe detuning $(\Delta_p/\gamma_{31})$ for different values of the control field $(\Omega_c/\gamma_{31})$ in Fig. 2a and Fig. 2b, respectively. From Fig. 2a it is amply clear that, in absence of the control field $(\Omega_c/\gamma_{31} = 0)$, the probe field is largely absorbed when it is at resonance $(\Delta_p/\gamma_{31} = 0)$. For a suitable control field $(\Omega_c/\gamma_{31} = 2)$, the absorption profile splits into two separate peaks, called Autler-Townes absorption doublet, which is the signature of formation of EIT window. The transparency window (TW) widens with the increase in the value of control field which is evident from the curve in figure for control field $(\Omega_c/\gamma_{31} = 2\ \&\ 4)$. Meanwhile, from Fig. 2b, it can be seen that, initially for $\Omega_c/\gamma_{31} = 0$, the profile of $Re(\chi^{(1)})$ possesses a negative slope as $\Delta_p/\gamma_{31}$ changes from $-ve$ to $+ve$ value within the TW. The dispersion profile has two regimes: normal and anomalous. For $Re(\chi^{(1)}) > 0$ i.e., in normal dispersion regime, we have $V_g < c$ that implies the probe field is slow comparison to the velocity of light. While for $Re(\chi^{(1)}) < 0$ i.e., in anomalous dispersion regime, one has $V_g > c$, hence, the probe field is fast comparison to the velocity of light. In the present case, when the control field is turned on and tuned to suitable values say at $\Omega_c/\gamma_{31} = 2$ and 4, the profile of $Re(\chi^{(1)})$ possesses positive slope at the centre of the TW. Thus the group velocity of the probe laser can be slowed down with negligible absorption. Thus far what we have discussed in this section is reported earlier and well understood[31,32].

We now proceed to study the third order susceptibilities. To begin with, in Fig. 3a we have demonstrated the variation of real part of third order nonlinear susceptibility $Re(\chi^{(3)})$ as functions of $\Delta_p/\gamma_{31}$ for different strength of control field ($\Omega_c/\gamma_{31}$) and zero detuning (i.e., $\Delta_c/\gamma_{31} = 0$). The third order susceptibility is quite large and possesses a single peak when no control field is applied. With the application of finite control field an additional peak in the susceptibility appears and the separation between these two peaks increases with the increase in the value of the control field. For clarity in understanding, the imaginary part of the linear susceptibility has been depicted at the bottom of the Fig. 3a. It is noteworthy to point out that the third order nonlinearity possesses large value within the EIT window, which will be subsequently exploited in the next section to generate supercontinuum.

In order to study the influence of control field detuning ($\Delta_c/\gamma_{31}$) on the third order nonlinearity, we have demonstrated in Fig. 3b the variation of $Re(\chi^{(3)})$ with $\Delta_p/\gamma_{31}$ for finite $\Delta_c/\gamma_{31}$. It is evident from figure that, in the absence of any control field detuning ($\Delta_c/\gamma_{31} = 0$), the variation of $Re(\chi^{(3)})$ shows antisymmetric behavior while for finite detuning ($\Delta_c/\gamma_{31} \neq 0$) it loses antisymmetric property. Therefore, the peak value of $Re(\chi^{(3)})$ can be shifted to any desired probe frequency by manipulating the Rabi frequency and detuning of the control field. At this stage a comparison of values of $\chi^{(3)}$ exhibited by different materials reported in the literature will be worthy, which is summarized in Table 1. From the values enlisted in the table, it is amply clear that the value of $\chi^{(3)}$ as identified in the present investigation is extremely large in comparison to that exhibited by other materials, particularly photonic crystal fibers which have been widely used for the generation of optical supercontinuum.

**Supercontinuum generation under EIT scheme**

In this section, we proceed to investigate optical supercontinuum generation exploiting the large nonlinearity exhibited within the transparency window due to EIT. The key to quality supercontinuum generation is large optical nonlinearity as well as low dispersion. Therefore, for the generation of supercontinuum at low power level, we need to select the value of the wavelength of the probe pulse such that it experiences negligible absorption, low dispersion and large nonlinearity. We therefore choose a probe detuning of $\Delta_p/\gamma_{31} = 0.8$, which corresponds to probe wavelength $\lambda_p = 9.963 \mu m$. At this probe wavelength, different parameters turn out to be $\chi^{(1)} = 0.1836 + i\,0.0434$, $Re(\chi^{(3)}) = 1.52 \times 10^{-13}$ m²/V², $\beta_2 = -1.08 \times 10^{-20} + i4.73 \times 10^{-21}$ s²/m, $W = 3.80 \times 10^{-21} + i8.97 \times 10^{-22}$ s²/m, $\gamma = 1.05 \times 10^7 \, W^{-1}m^{-1}$, and the dispersion $D = -\frac{2\pi c}{\lambda_p^2}\beta_2 = 0.21 \times 10^3 \, ps.nm^{-1}m^{-1}$. In Fig. 4a and 4b, we have demonstrated the variations of $\beta_2$ and $D$ with $\Delta_p/\gamma_{31}$ and wavelength respectively, while Fig. 4c demonstrates the variation of $Im(\chi^{(1)})$ with $\Delta_p/\gamma_{31}$. The probe wavelength has been marked with a broken vertical line in each of these subplots. Note that without the application of the control field the value of $Im(\chi^{(1)})$ is 0.5323, while it reduces to 0.0434 when the control field is tuned to $\Omega_c/\gamma_{31} = 4$ and detuning to $\Delta_c/\gamma_{31} = -3$. Since imaginary parts of $\beta_2$ and $W$ are small in comparison to their corresponding real counterpart, we can neglect them without any loss of generality and proceed to study supercontinuum generation using equation (12).

In order to investigate SC generation adopting numerical simulation, we have launched 100fs unchirped sech pulses at 9.963 μm wavelength in a 1.374 μm long MQW system which is composed of 30 well periods. Peak power of these pulses is 2.0 W. The spectral and temporal evolution of the input pulse has been displayed for different length of MQW system in Fig. 5a-(i) and (ii), respectively. It is amply clear that significant spectral broadening has been achieved at the end of the MQW system. The spectral broadening is attributed to self phase modulation (SPM) and modulation instability. Several dips in the spectra, as evident from Fig. 5a-(i), are due to modulation instability. The launched pulse undergoes splitting due to modulation instability at the end of the MQW. In order to get additional information about spectral and temporal broadening dynamics and also to understand the influence of pulse peak power on the supercontinuum, we have injected 100fs pulses of different peak power, particularly 1.0 W, 1.5 W and 2.0 W into the MQW and captured generated supercontinuum. The spectral as well as the temporal profiles have been depicted in Fig. 5b using logarithmic density scale truncated at $-40dB$ relative to the maximum value. For low peak power say at 1.0 W, the injected pulse, while propagating through the MQW, undergoes temporal compression twice, one at a distance of 0.3 μm and other at 1.0 μm inside the MQW system. Consequently, the spectral widths at these two distances are widest. Along the length of the well the spectrum undergoes periodic expansion and compression, frequency of which increases with the increase in peak power of injected pulses. Linking to the explanation stated above for Fig. 5a, we present an insight investigation on the density plots for 2.0 W peak power depicted in Fig. 5b (top panel). Initially the spectral broadening for 2.0 W peak power is attributed to nonlinear self-phase modulation. After sufficient spectral broadening and temporal compression the pulse tries to broaden in temporal domain due to solitonic effect. As the pulse propagates, its temporal width breadths and accordingly in the spectral domain the spectra of the pulse also breadths, i.e., it first broadens then compresses, again broadens and compresses and so on. Gradually the modulation instability picks up and eventually the pulse splits into two. Trajectory of the pulse, as evident from the temporal profile, is straight indicating group velocity of the soliton remains same. In addition, large wavelength side of the spectra broadens more in comparison to the lower wavelength side. This is due to efficient dispersive wave generation in the longer wavelength. The final broadening of the pulse may be attributed to SPM and modulation instability. Several dips in the final SC spectra obtained at the end of the MQW are due to modulation instability. Fig. 5c demonstrates the spectral and temporal profiles at the end of the MQW system for pulses of three different peak powers 1.0 W, 1.5 W and 2.0 W. For pulses of 1.0 W peak power, the broadened spectrum at the end of the fiber is simple and devoid of complex internal structure. With the increase in power the broadened spectra develops dip in the structure, which is a signature of modulation instability. Finally, multiple dips appear in the central part of the broadened spectra when peak power of input pulses reach 2.0W. The initial pulses splits into two separate pulses of unequal shape which is evident from panel (ii) of Fig. 5c.

**Conclusion:**

We have investigated broadband mid-infrared supercontinuum generation at very low power in semiconductor multiple quantum well (MQW) systems facilitated by electromagnetically induced transparency. Using numerical simulation, we have demonstrated broadband supercontinuum generation resulting in due to the launching of 100 femto-seconds pulses of peak power close to a Watt in the electromagnetically induced transparency window of a 30 period $1.374\ \mu m$ long MQW system. The supercontinuum spectra, attributed to self phase modulation and modulation instability. The central part of the spectra is dominated by several dips and broadening of the far infra-red part of the spectra is more in comparison to the infra-red portion. Key advantage of the proposed scheme is that the supercontinuum source could be easily integrated with other semiconductor devices.

**Acknowledgement**

This work is supported by the University Grants Commission, Bahadur Shah Zafar Marg, New Delhi, through Major Project, F. No. 41-909/2012 (SR). Authors NB and SK thank UGC for financial support. The present investigation is also supported by Qatar National Research Fund (QNRF) trough the proposal NPRP 6-021-1-005. MB and SK acknowledge the support of QNRF with thanks.

**Author contributions**
All authors contributed equally.

**Competing financial interest**
The authors declare no competing financial interests.

**Figure Captions:**

Figure 1: (a) Schematic of the band structure of a single period of the multiple quantum well. Each InGaAs/AlInAs/InGaAs well has thickness of 9.8 nm which is covered by 36 nm AlInAs barriers.
(b) Energy level diagram of the quantum well. Arrows represent the ladder type excitation scheme.

Figure 2: Variation of (a) imaginary and (b) real parts of $\chi^{(1)}$ as a function of $\Delta_p/\gamma_{31}$ for three different values of normalized control detuning. $\Omega_c/\gamma_{31} = 0$ (dash), $\Omega_c/\gamma_{31} = 2$ (dot-dash) and $\Omega_c/\gamma_{31} = 4$ (solid).

Figure 3: (a) Variation of $Re(\chi^{(3)})$ (top-3 panels) and imaginary part of $\chi^{(1)}$ (bottom panel) as a function of $\Delta_p/\gamma_{31}$ for three different values of control Rabi frequency. Control field detuning $\Delta_c/\gamma_{31} = 0$. In both cases, $(i)\,\Omega_c/\gamma_{31} = 0$, $(ii)\,\Omega_c/\gamma_{31} = 2$ and $(iii)\,\Omega_c/\gamma_{31} = 4$. $Im(\chi^{(1)})$ has been depicted to show that nonlinearity is large even within the EIT window.
(b) Variation of $Re(\chi^{(3)})$ (top-3 panels) and imaginary part of $\chi^{(1)}$ (bottom panel) as a function of $\Delta_p/\gamma_{31}$ for three different values of control detuning. Rabi frequency of the control field $\Omega_c/\gamma_{31} = 4$. In both cases, $(i)\,\Delta_c/\gamma_{31} = 3$, $(ii)\,\Delta_c/\gamma_{31} = 0$ and $(iii)\,\Delta_c/\gamma_{31} = -3$. $Im(\chi^{(1)})$ has been depicted to confirm the existence of large nonlinearity within TW.

Figure 4: Variations of $\beta_2$ and chromatic dispersion $D$ with $\Delta_p/\gamma_{31}$ and wavelength, have been depicted in panels (a) and (b), respectively. In the bottom panel $Im(\chi^{(1)})$ has been depicted to identify the pump wavelength that has been marked by vertical broken line. At pump wavelength $9.963\,\mu m$, dispersion $D$ is anomalous. Without the application of control field $Im(\chi^{(1)}) = 0.5323$, while it reduces to $0.0434$ when control field is applied.

Figure 5: (a) Spectral and temporal profile of the supercontinuum obtained due to input pulse of 100 fs duration and 2.0 W peak power.
(b) Density plot of the spectral and temporal evolution of 100 fs input pulse with three different peak power 1.0 W, 1.5 W and 2.0 W.
(c) Spectral and temporal profiles of the pulse at the end of the MQW for three different peak powers: 1.0 W, 1.5 W and 2.0 W. In all cases, (i) spectral profiles and (ii) temporal profiles.

**Table:**

Table 1: Comparison of values of $\chi^{(3)}$ for different materials, QW nanostructures and the QW structure considered in the present investigation.

**Figures:**

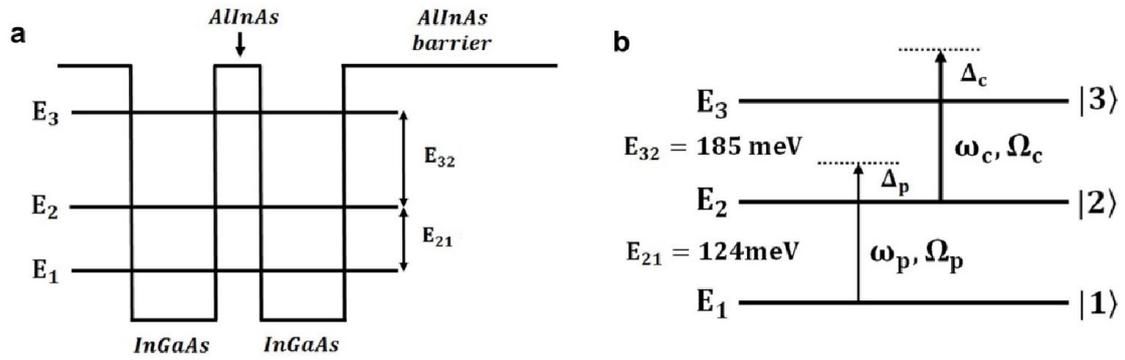

Figure 1: (a) Schematic of the band structure of a single period of the multiple quantum well. Each InGaAs/AlInAs/InGaAs well has thickness of 9.8 nm which is covered by 36 nm AlInAs barriers.
(b) Energy level diagram of the quantum well. Arrows represent the ladder type excitation scheme.

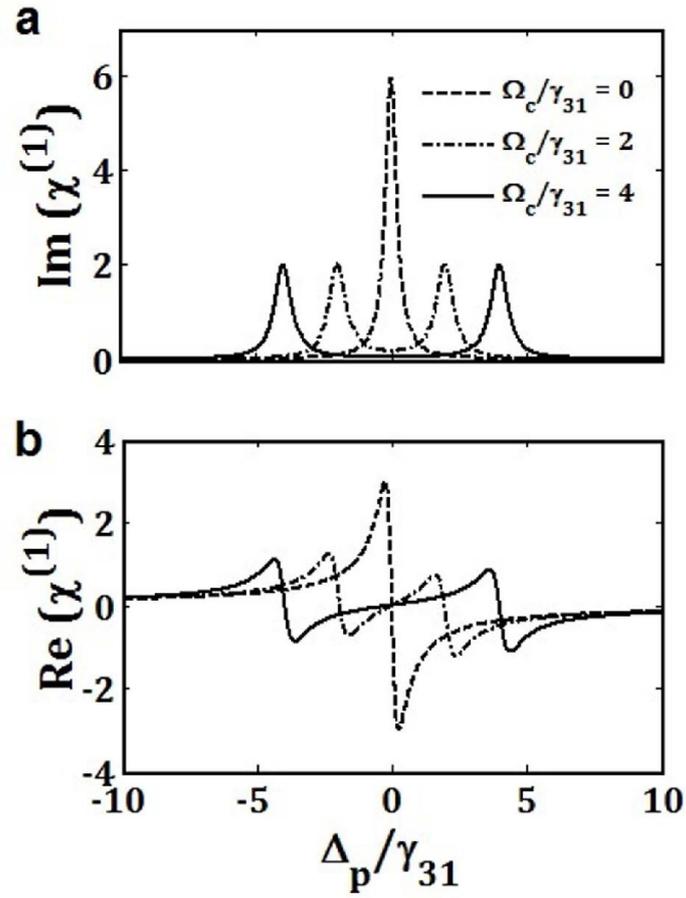

Figure 2: Variation of (a) imaginary and (b) real parts of $\chi^{(1)}$ as a function of $\Delta_p/\gamma_{31}$ for three different values of normalized control detuning. $\Omega_c/\gamma_{31} = 0$ (dash), $\Omega_c/\gamma_{31} = 2$ (dot-dash) and $\Omega_c/\gamma_{31} = 4$ (solid).

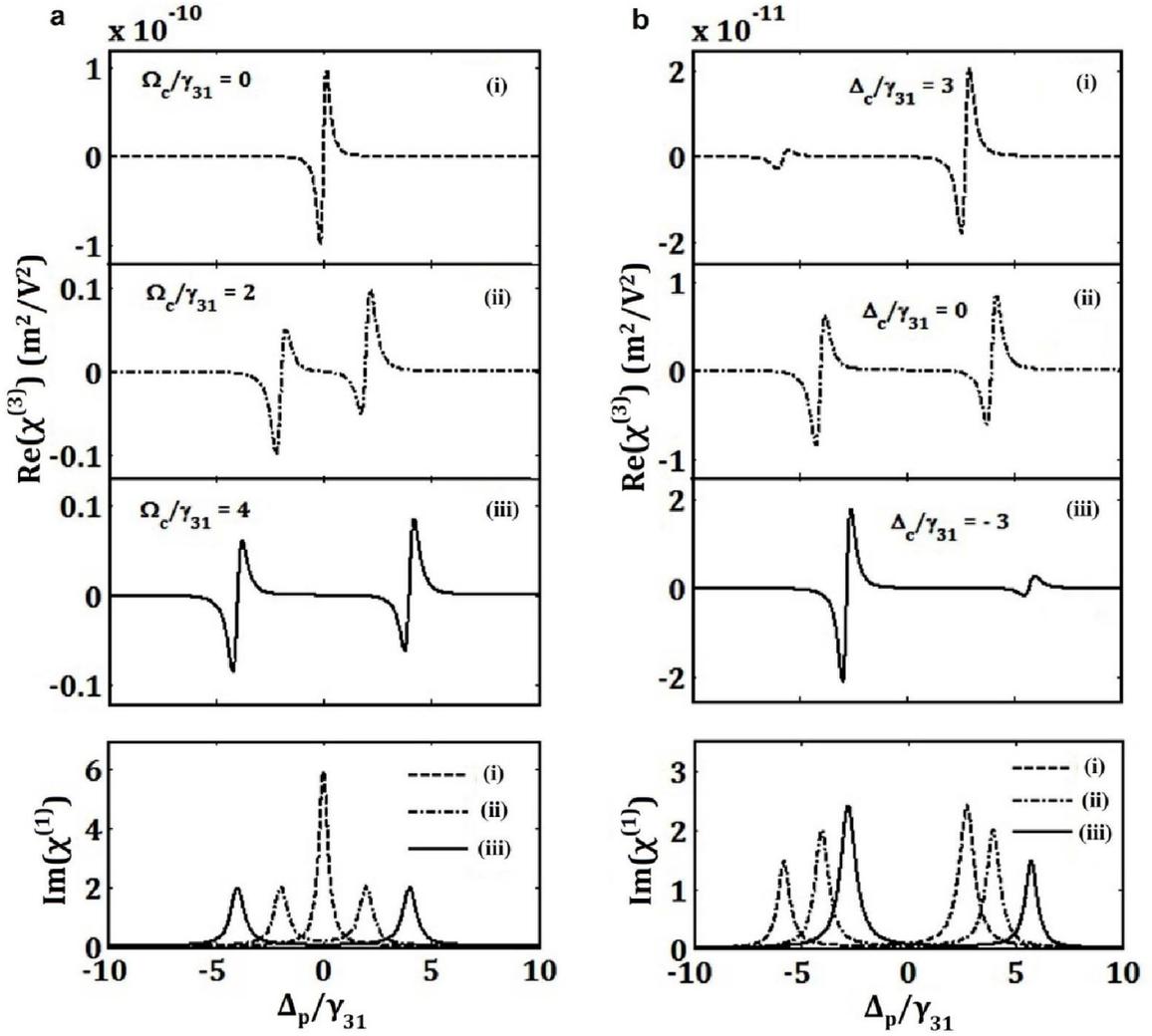

Figure 3: (a) Variation of $Re\chi^{(3)}$ (top-3 panels) and imaginary part of $\chi^{(1)}$ (bottom panel) as a function of $\Delta_p/\gamma_{31}$ for three different values of control Rabi frequency. Control field detuning $\Delta_c/\gamma_{31} = 0$. In both cases, $(i)\,\Omega_c/\gamma_{31} = 0$, $(ii)\,\Omega_c/\gamma_{31} = 2$ and $(iii)\,\Omega_c/\gamma_{31} = 4$. $Im(\chi^{(1)})$ has been depicted to show that nonlinearity is large even within the EIT window.

(b) Variation of $Re\chi^{(3)}$ (top-3 panels) and imaginary part of $\chi^{(1)}$ (bottom panel) as a function of $\Delta_p/\gamma_{31}$ for three different values of control detuning. Rabi frequency of the control field $\Omega_c/\gamma_{31} = 4$. In both cases, $(i)\,\Delta_c/\gamma_{31} = 3$, $(ii)\,\Delta_c/\gamma_{31} = 0$ and $(iii)\,\Delta_c/\gamma_{31} = -3$. $Im(\chi^{(1)})$ has been depicted to confirm the existence of large nonlinearity within TW.

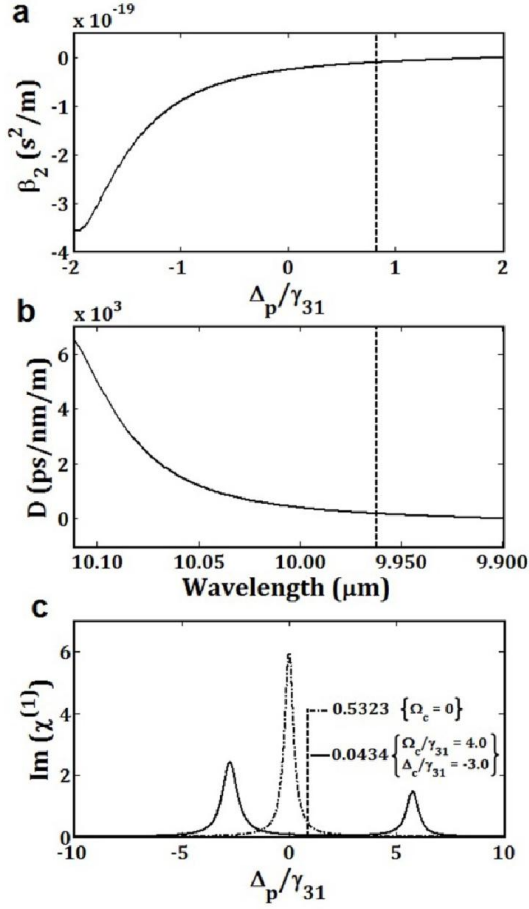

Figure 4: Variations of $\beta_2$ and chromatic dispersion $D$ with $\Delta_p/\gamma_{31}$ and wavelength, have been depicted in panels (a) and (b), respectively. In the bottom panel $Im(\chi^{(1)})$ has been depicted to identify the pump wavelength that has been marked by vertical broken line. At pump wavelength 9.963 $\mu m$, dispersion $D$ is anomalous. Without the application of control field $Im(\chi^{(1)}) = 0.5323$, while it reduces to 0.0434 when control field is applied.

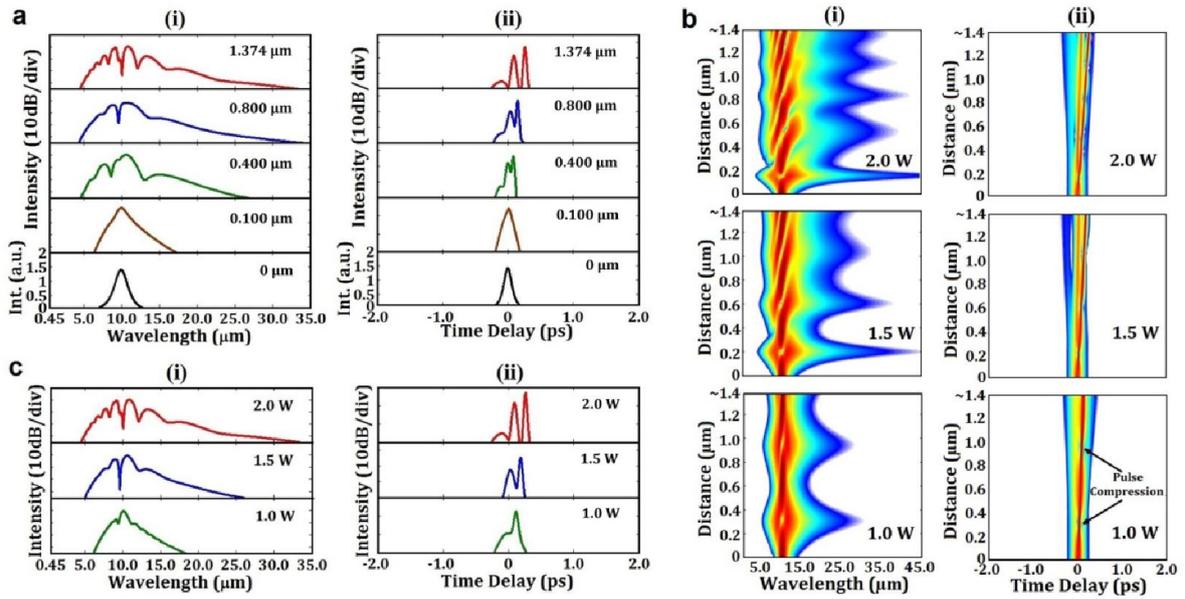

Figure 5: (a) Spectral and temporal profiles of the supercontinuum obtained due to input pulse of 100 fs duration and 2.0 W peak power.
(b) Density plot of the spectral and temporal evolution of 100 fs input pulse with three different peak power 1.0 W, 1.5 W and 2.0 W.
(c) Spectral and temporal profiles of the pulse at the end of the MQW for three different peak powers: 1.0 W, 1.5 W and 2.0 W. In all cases, (i) spectral profiles and (ii) temporal profiles.

Table 1: Comparison of values of $\chi^{(3)}$ for different materials, QW nanostructures and the QW structure considered in the present investigation.

| Materials | $\chi^{(3)} (m^2/V^2)$ | Wavelength (µm) | Ref.s |
|---|---|---|---|
| Fused Silica | $1.9 \times 10^{-22}$ | 0.800 | 33 |
| PBG-08 PCF | $2.7 \times 10^{-21}$ | 0.900 | 33 |
| Silicate N-F2 PCF | $1.4 \times 10^{-21}$ | 1.000 | 33 |
| Semiconductor Doped Glass fiber | $4.5 \times 10^{-19}$ | 0.740 | 34 |
| P-toluene sulphonate (PTS) | $3.7 \times 10^{-18}$ | 1.060 | 35 |
| **Quantum Wells** | | | |
| InGaAs/AlAs/AlAsSb | $5.8 \times 10^{-17}$ | 1.550 | 36 |
| GaN/AlN | $2.2 \times 10^{-16}$ | 1.550 | 37 |
| Si doped GaN–AlN | $2.2 \times 10^{-15}$ | 1.500 | 38 |
| GaAs/AlInAs | $1.5 \times 10^{-13}$ | 9.963 | Present Paper |